\documentclass[proof]{WileyASNA-v1}

\articletype{Article Type}%

\received{<day> <Month>, <year>}
\revised{<day> <Month>, <year>}
\accepted{<day> <Month>, <year>}

\usepackage{moreverb}
\usepackage{graphicx}

\begin{document}

\title{Jet propagation through inhomogeneous media and shock ionization}

\author[1,2]{Manel Perucho*}

\author[1]{Jose L\'opez-Miralles}

\author[3]{Victoria Reynaldi}

\author[4,5]{\'Alvaro Labiano}

\authormark{Perucho, L\'opez-Miralles, Reynaldi, Labiano}

\address[1]{\orgdiv{Departament d'Astronomia i Astrof\'{\i}sica}, \orgname{Universitat de Val\`encia}, \orgaddress{\state{Val\`encia}, \country{Spain}}}

\address[2]{\orgdiv{Observatori Astron\`omic}, \orgname{Universitat de Val\`encia}, \orgaddress{\state{Val\`encia}, \country{Spain}}}

\address[3]{\orgdiv{Facultad de Ciencias Astronómicas y Geofísicas}, \orgname{Universidad Nacional de La Plata}, \orgaddress{\state{La Plata}, \country{Argentina}}}

\address[4]{\orgdiv{Centro de Astrobiolog\'{\i}a}, \orgname{CSIC-INTA}, \orgaddress{\state{Madrid}, \country{Spain}}}

\address[5]{\orgdiv{Telespazio UK}, \orgname{for the European Space Agency (ESA)}, \orgaddress{\state{ESAC, Villanueva de la Ca\~nada, Madrid}, \country{Spain}}}

\corres{*Manel Perucho Pla, C/ Dr. Moliner 50, 46100, Burjassot, Valencian Country, Spain. \email{manel.perucho@valencia.edu}}


\abstract[Abstract]{In this contribution we present the first numerical simulations of a relativistic outflow propagating through the inner hundreds of parsecs of its host galaxy, including atomic and ionised hydrogen, and the cooling effects of ionisation. Our results are preliminary, but we observe efficient shock ionization of atomic hydrogen in interstellar clouds. The mean density of the interstellar medium in these initial simulations is lower than that expected in typical galaxies, which makes cooling times longer and thus no recombination is observed inside the shocked region. The velocities achieved by the shocked gas in the simulations are in agreement with observational results, although with a wide spectrum of values.}


\jnlcitation{\cname{
\author{Perucho, M.}, 
\author{L\'opez-Miralles, J.}, 
\author{Reynaldi, V.}, and 
\author{Labiano, A.}} (\cyear{2021}), 
\ctitle{Jet propagation through inhomogenous media and shock ionization}, \cjournal{Astronomische Nachrichten}, \cvol{2021;XXX}.}

\maketitle


\section{Introduction}\label{sec1}

The jets are triggered around supermassive black holes in Active Galactic Nuclei (AGN) via the extraction of rotational energy from the black hole-accretion disk systems that sit at the core of those galaxies \citep{1977MNRAS.179..433B,2019ApJ...875L...1E}. Jets become relativistic and super(magneto)sonic beyond the acceleration region, thus triggering shocks in the galactic interstellar medium (ISM) during the first stages of propagation through the host galaxy. 

The role of shocks is certainly relevant for the subsequent evolution of the host galaxy. On the one hand, shocks heat and ionize the ISM, providing it with momentum and triggering outflows of ISM gas \citep[e.g.,][and references therein]{2016A&A...593A..30M,2018A&ARv..26....4M,2021A&A...647A..63S}. These outflows could eventually extract large amounts of ISM gas from the galaxy, with the consequent quenching of star formation \citep[see, e.g.,][]{2011ApJ...743...42P,2014MNRAS.445.1462P}. On the other hand, there is debate about the possibility that the compression induced by shocks could cause a burst of star formation, although heated up ISM would increase the turbulence in its molecular gas, which would prevent the collapse of the star foming cloud \citep[see the discussion in, e.g.][]{2012ApJ...757..136W}.

During the last years, a number of numerical simulations have tackled the evolution of jets through the two-phase, inhomogeneous media expected within the inner kiloparsecs of host galaxies \citep{2011ApJ...728...29W,2012ApJ...757..136W,2016MNRAS.461..967M,2018MNRAS.475.3493B,2019MNRAS.489.4944Z}. These works have shown that jet evolution can be strongly modified by the presence of local, dense inhomogeneities and these are shocked and destroyed by jets in time-scales that could allow star formation. Although the physical conditions of shocked clouds are complex and more detailed, specific studies should be performed.

\begin{table*}[t]%
\caption{Simulation parameters. $v$ is velocity, $\rho_j/\bar{\rho_a}$ is the density ratio, $T$ is temperature, $L_j$ is jet power, and $n$ is the number density. \label{tab1}}
\centering
\begin{tabular*}{500pt}{@{\extracolsep\fill}lccccccc{c}{c}{c}c@{\extracolsep\fill}}
\toprule
&\multicolumn{4}{@{}c@{}}{\textbf{Jet properties}} & \multicolumn{3}{@{}c@{}}{\textbf{ISM properties}} \\\cmidrule{2-5}\cmidrule{6-8}
\textbf{Model} & $v\,(c)$  & $\rho_j / \bar{\rho_a} $  & $T\, {\rm (K)}$  & $L_j$ (${\rm erg\, s^{-1}}$) &
$\bar{n_a} (\rm cm^{-3})^\tnote{*}$   & $\rho\,({\rm m_p\, cm^{-3}})^{**}$ & T (K)$^{**}$\\
\midrule
J46a & 0.98  & $10^{-4}$  & $2 \times 10^{12}$  & $10^{46}$ & 0.1 & $10^{-3}\,-\,10^{1}$ & $10^3-10^8$ \\
J44a & 0.98  & $10^{-5}$  & $2 \times 10^{11}$  & $10^{44}$ &  0.1 & $10^{-3}\,-\,10^{1}$& $10^3-10^8$ \\
J46b & 0.98  & $10^{-5}$  & $2 \times 10^{12}$  & $10^{46}$ &  1 &$10^{-2}\,-\,10^{2}$ & $10^3-10^8$\\
\bottomrule
\end{tabular*}
\begin{tablenotes}
\item[* ] Peak of the log-normal number density distribution.
\item[** ] These numbers indicate the ranges of values with largest occurrence in the grid, i.e., smaller or larger values occur with a very small probability and include a negligible number of cells.  
\end{tablenotes}
\end{table*}

In this work, we present our first simulations to study the role of jets within their host galaxies. The novelty of these numerical experiments is the addition of atomic hydrogen as a species to our simulations, along with the effects of cooling and recombination of hydrogen in terms of radiative losses. 


\begin{figure}
\includegraphics[width=\columnwidth]{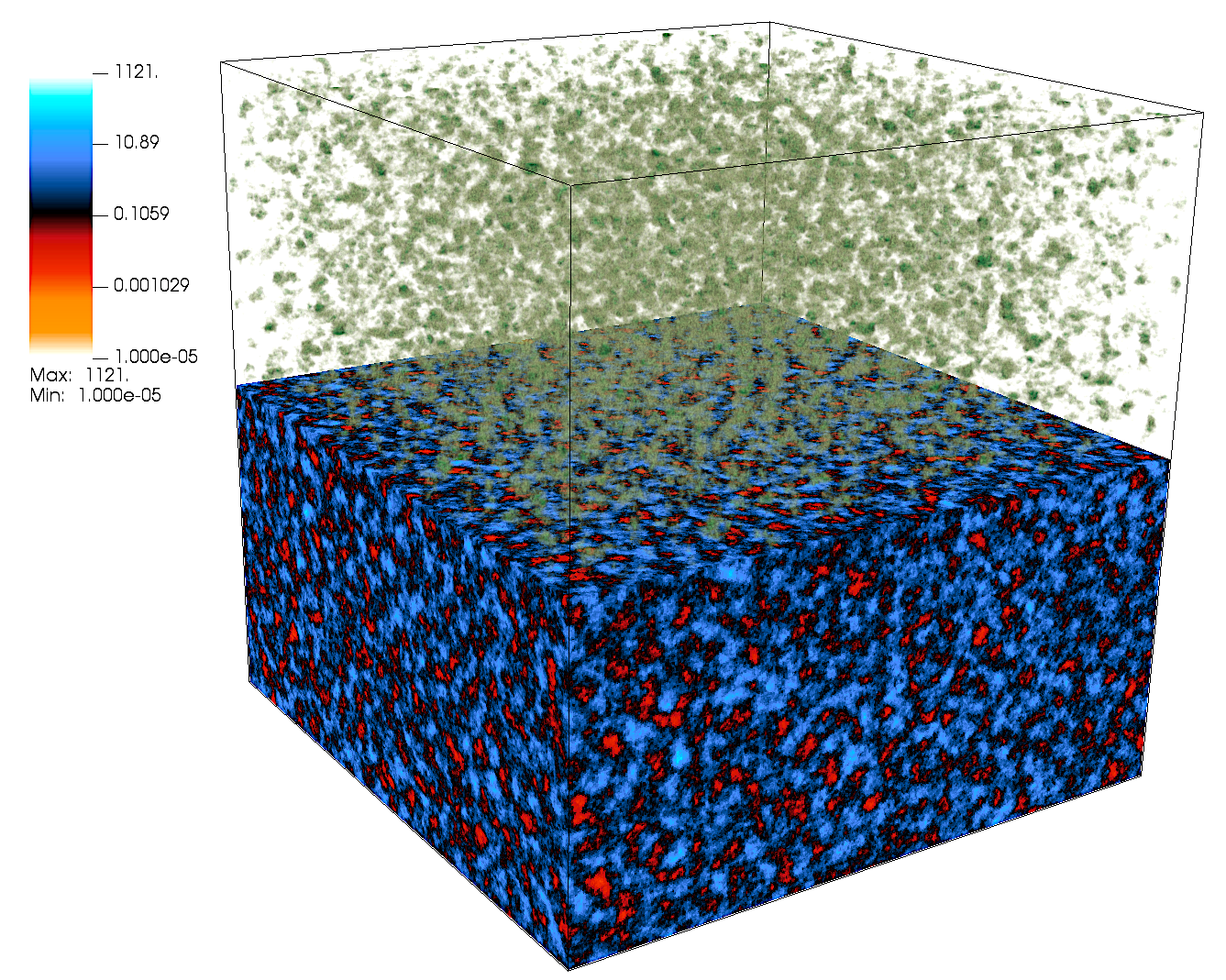}
\caption{Initial setup, including a slice of density (bottom) and atomic hydrogen clouds (top).  \label{fig3}}
\end{figure}

\section{Numerical simulations}\label{sims}
With the aim to run numerical simulations that account for ionisation of hydrogen, we have modified the code \emph{Ratpenat} \citep{2010A&A...519A..41P,2019MNRAS.482.3718P} to include atomic hydrogen and ionization terms from \citet{2015A&A...580A.110V}. We have chosen to implement the non-equilibrium conditions, because of the low densities in both AGN jets and the ISM. Therefore, ionisaton plays a role as a source of 1) energetic losses, which means a source term to include the collisional ionisation cooling and the radiative recombination in the energy conservation equation, and 2) the particle number densities, i.e., it appears as a source term in the conservation equation of the different species. We have kept the Synge Equation of State (EoS) for the jet flow, composed of pairs with high internal energies. The selection of the EoS for each cell is based on the presence or not of atomic hydrogen (see Perucho et al., in preparation, for details on the code). In contrast to the work by \citet{2015A&A...580A.110V}, we have not included neither atomic helium nor molecular hydrogen in this first step of our project. The energy density of the gas is thus determined by the translational energy of hydrogen alone \citep[see][]{2015A&A...580A.110V} and the thermodynamical relations become trivial.

Despite the lack of analytic solutions for relativistic shock propagation in combination with hydrogen ionisation\footnote{\citet{2015A&A...580A.110V} limit their work to classical flows.}, we have run a number of tests that have been successfully passed by the code (Perucho et al., in preparation), and compare well with those presented in \citet{2015A&A...580A.110V}. 

The simulations were run in 512 cores at Tirant, the local supercomputer at the University of Val\`encia. The grid size is $512\times512\times512$~cells, with a resolution of 6 cells/$R_{\rm j}$, and $R_{\rm j} = 6\,{\rm pc}$ at injection. This means that the simulation follows the jet evolution along 512~pc, from the point where the jet radius is the selected one (i.e., $\sim 60$~pc from the active nucleus). 

The initial 3D density distribution was set by means of the PyFC fractal code \citep{2011ApJ...728...29W}. This algorithm implements an iterative process such that density inhomogeneities follow a log-normal single-point statistics in space ($\mu=1.0,\sigma^2=5.0$). The fractal structure of the scalar field is achieved multiplying the Fourier-transformed distribution by a Kolmogorov power-law with spectral index $\beta=-5/3$. The scale of the largest cloud in the box is approximately 16\,pc, which corresponds to a minimum sampling wave-number, $k_\mathrm{min}=17$. The medium is set in pressure equilibrium to a given value ($P_{\rm ISM}= 3.4\times10^{-10} - 3.4\times10^{-11}\,{\rm erg\,cm^{-3}}$), depending on the mean ambient density, by adapting temperatures. Once temperatures are fixed, we use Saha equilibrium equations to set an initial value for hydrogen density. The medium is obviously out from the equilibrium required by Saha, and is thus dynamic, in the sense that the hydrogen density changes with time due to cooling. The numerical box is assumed to be optically thin, so all the energy lost via cooling is assumed to be radiated away.

We have run three numerical simulations: J46a, J46b and J44a. Simulations J46 share the same jet properties and gas distribution, but the ISM mean density is scaled up by an order of magnitude in simulation J46b with respect to J46a. Simulation J44a shares the ambient medium properties with J46a, but the jet injected in J44a is two orders of magnitude lower in power with respect to J46a. Table \ref{tab1} shows the parameters used in these simulations. The jets are all purely leptonic at injection. A tracer has been set to 1 at the densest and coldest regions and 0 for the rest of the grid, with the aim to track the original material in ISM clouds. Figure~\ref{fig3} shows three-dimensional contours of atomic hydrogen density in the grid prior to jet injection. We have used LLNL VisIt \citep{Childs} to produce the figures.

\section{Preliminary Results} \label{res}

\begin{figure}[t!]


\includegraphics[width=\columnwidth]{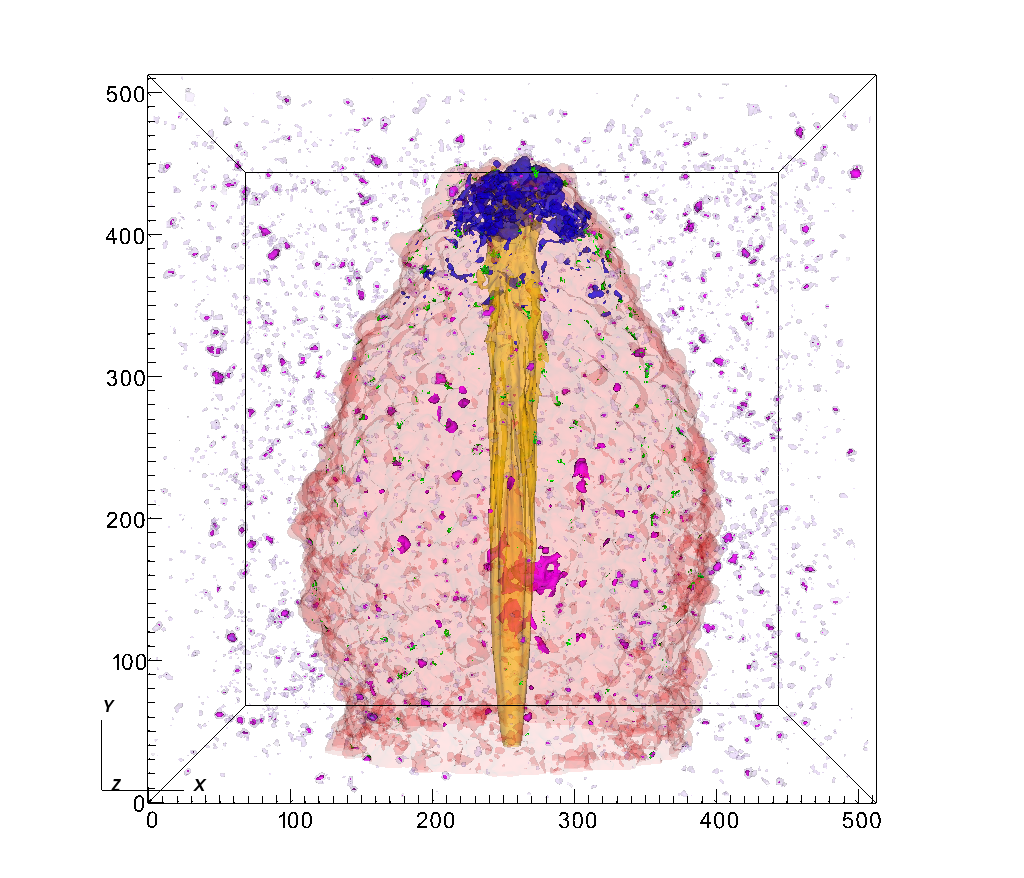}
\caption{Last frame of simulation J46b, within a 512~pc side box. Pressure contours are red at the bow-shock ($P=10^{-4}\, c^2$), and dark blue at the hot-spot ($P=6\times 10^{-3}\, c^2$); leptonic fraction contour is orange ($x_e=0.5$); tracer contour is pink ($f=0.5$), and those showing the atomic hydrogen density inside and outside the bow shock, are green/yellow ($n_H=0.1$ and $n_H=1\,{\rm cm^{-3}}$) and purple ($10\,{\rm cm^{-3}}$), respectively. \label{fig4}}

\end{figure}

Figure~\ref{fig4} shows, for the last snapshot of simulation J46b, the projected 3D distribution of pressure contours (red and dark blue) indicating the position of the bow-shock and hot-spot; a leptonic fraction contour (orange) indicating the jet surface; a tracer contour (pink) indicating the location of material originally sitting in the densest and coldest regions that has been shock-ionized; and contours showing the atomic hydrogen density in two different sets: the blueish contours indicate hydrogen outside the bow shock, whereas the green/yellow contours indicate hydrogen inside the red surface. Altogether, the image shows how the inhomogeneous density of the ISM forces an irregular structure in the bow-shock. The distortion and expansion that the passage of the shock and, mainly, close passage by the jet, forces in the cloud material, as shown by the pink spots at the centre of the image, is remarkable. Images not shown here provide evidence of strong mixing and chaotic motions inside the shocked region, causing an efficient mixing between the original cloud gas and the dilute, hotter ISM. The original gas inside the clouds is thus scattered through larger volumes, as revealed by the image.

The jet in simulation J46a crossed the grid in $\sim 5100$~years, which implies an advance velocity of $\simeq 0.3\,c$. In the case of J46b, the simulation time is $\sim 6700$~years (advance speed $\simeq 0.25\,c$), whereas in the case of J44a (not shown), it is $\sim 9000$~years (advance speed $\simeq 0.18\,c$). These advance velocities are just slightly above those observed by \citet{1998A&A...337...69O,2003PASA...20...69P} for Compact Symmetric Objects, probably due to the dilute media used in these initial simulations. 

All the green/yellow spots lie close to the bow-shock discontinuity, showing that all the atomic hydrogen gets completely ionized as the shock crosses the cloud. This result has been confirmed by histograms (not shown here) of the number of cells with atomic hydrogen embedded inside a given pressure. These histograms show that the cells containing atomic hydrogen rapidly tend to zero with increasing pressure, close to the isobaric surface shown in red in the image, as temperatures reach $10^4 - 10^5$~K. The same behaviour found in J46a is seen in both J46b and J44a: atomic hydrogen seems to be ionized as the shock crosses the clouds. 



The temperatures measured inside the shocked regions are extremely high ($\sim10^{10}-10^{11}$~K), which makes it impossible for the ionised gas to cool and recombine within the simulation time-scales. The heating is due to both the shock and mixing with shocked jet material, and the high temperatures in the relatively dilute space between cold clouds (as forced by the large span of ISM density values and the imposed pressure equilibrium in the ambient medium, see Table~\ref{tab1}). These initially already high temperatures get increased by the shock, and facilitate reaching the observed values within the shocked region via mixing.

\begin{figure}
\begin{center}
\includegraphics[trim=4.5cm 4cm 0 0,width=0.9\columnwidth]{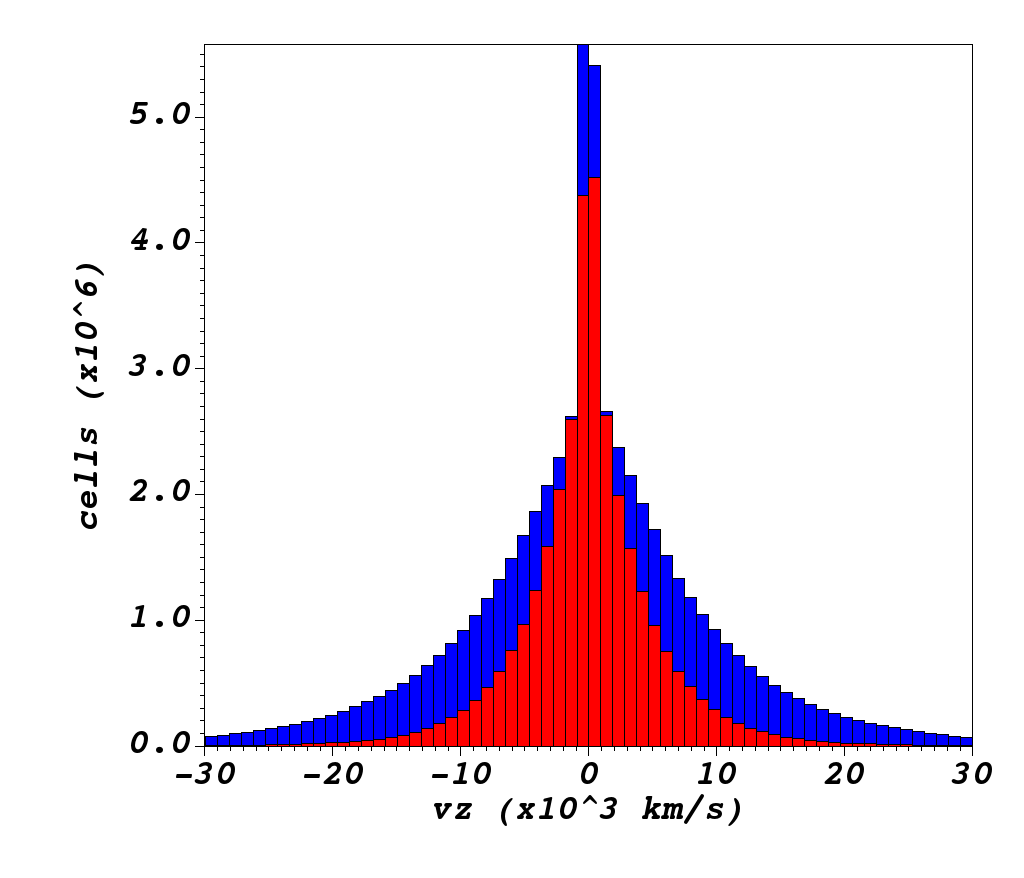}
\end{center}
\caption{Histogram of the number of cells ($\times 10^6$) within the bow-shock with a given $v_z$ ($\times 10^3\,{\rm km/s}$), weighted with density, thus revealing transversal velocities with respect to the jet flow direction of the densest shocked regions. Blue columns stand for simulation J46a and red for J46b. \label{fig5}}
\end{figure}

We have computed histograms of transversal (see, e.g., Fig.~\ref{fig5}) and longitudinal velocities (with respect to the jet direction) in cells inside the shocked region. The count has been weighted with density in order to focus on the cells with a larger densities (and original cloud material). The results show that, in J46b, where the jet propagates through a denser medium, the most common velocities are $\leq 5000~{\rm km/s}$ (also in the case of J44a, not shown), with the largest columns indicating velocities $\sim10^{2-3}$~km/s, whereas a significant number of cells reach up to $10^4$~km/s in J46a. The longitudinal velocity distributions also show the trend towards narrower wings (smaller velocities) in J46b and J44a.


The axial velocity distributions show fairly symmetric profiles around zero, which implies that the shocked ISM is involved in both forward motions and backflow towards the galactic nucleus, probably when mixed with the shocked jet gas. Altogether, this reveals chaotic motions, which translates in the ideal scenario for turbulence. 


These results show that, in the lower velocity limit, our simulations are capable of reproducing the observed kinematics in Compact Steep Spectrum (CSS) sources such as those reported in \citet{labiano05} and \citet{reynaldi16} through ionised gas observations. Furthermore, the structure and distribution of the ionized clouds (pink contours in Fig. 2) is consistent with the ionised fast clouds observed in CSS  \citep[see also e.g.,][]{Shih13,Holt11,ODea02}. A refinement on both the ISM initial conditions and jet properties are mandatory in order reach the velocity ranges with which the ionized gas is observed in CSS sources.







\section*{Acknowledgments}

\footnotesize{Computer simulations have been carried out in the the supercomputer Tirant, at the Servei d'Inform\`atica de la Universitat de Val\`encia.  This work has been supported by the Spanish Ministry of Science through Grants PID2019-105510GB-C31, PID2019-107427GB-C33, PID2019-106280GB-I00 (MCIU/AEI/FEDER,UE), and from the Generalitat Valenciana through grant PROMETEU/2019/071. A.L. acknowledges the support from Comunidad de Madrid through the Atracción de Talento Investigador Grant 2017-T1/TIC-5213. JLM acknowledges the support of a fellowship from ‘La Caixa’ Foundation (ID 100010434). The fellowship code is LCF/BQ/DR19/11740030. V.R. acknowledges the support from Universidad Nacional de La Plata through grant 11/G153.}
  
\nocite{*}
\bibliography{Wiley-ASNA}

\begin{thebibliography}{}

\bibitem [\protect \citeauthoryear {%
Bettonvil%
, Sutterlin%
, Hammerschlag%
, Rutten%
\BCBL {}\ \BBA {} Stix%
}{%
Bettonvil%
\ \protect \BOthers {.}}{%
{\protect \APACyear {2003}}%
}]{%
Bettonvil2003}
\APACinsertmetastar {%
Bettonvil2003}%
\begin{APACrefauthors}%
Bettonvil, F\BPBI C.%
, Sutterlin, P.%
, Hammerschlag, R\BPBI H.%
, Rutten, R\BPBI J.%
\BCBL {}\ \BBA {} Stix, M.%
\end{APACrefauthors}%
\unskip\
\newblock
\APACrefYearMonthDay{2003}{}{},
\newblock
{\BBOQ}\APACrefatitle {Proc. SPIE Conf. Ser.} {Proc. SPIE Conf. Ser.}{\BBCQ}
\newblock
\BIn{} \BVOL\ 4853, \BPG~306.
\PrintBackRefs{\CurrentBib}

\bibitem [\protect \citeauthoryear {%
Bland-Hawthorn%
, van Breugel%
\BCBL {}\ \BBA {} Gillingham%
}{%
Bland-Hawthorn%
\ \protect \BOthers {.}}{%
{\protect \APACyear {2001}}%
}]{%
Bland2001}
\APACinsertmetastar {%
Bland2001}%
\begin{APACrefauthors}%
Bland-Hawthorn, J.%
, van Breugel, W.%
\BCBL {}\ \BBA {} Gillingham, P\BPBI R.%
\end{APACrefauthors}%
\unskip\
\newblock
\APACrefYearMonthDay{2001}{}{},
\newblock
\unskip
\newblock
\APACjournalVolNumPages{ApJ}{563}{}{611}.
\PrintBackRefs{\CurrentBib}

\bibitem [\protect \citeauthoryear {%
Kosugi%
\ \BBA {} Gillingham%
}{%
Kosugi%
\ \BBA {} Gillingham%
}{%
{\protect \APACyear {2007}}%
}]{%
Kosugi2007}
\APACinsertmetastar {%
Kosugi2007}%
\begin{APACrefauthors}%
Kosugi, T.%
\BCBT {}\ \BBA {} Gillingham, R\BPBI H.%
\end{APACrefauthors}%
\unskip\
\newblock
\APACrefYearMonthDay{2007}{}{},
\newblock
\unskip
\newblock
\APACjournalVolNumPages{Sol. Phys.}{243}{}{3}.
\PrintBackRefs{\CurrentBib}

\bibitem [\protect \citeauthoryear {%
Kosugi%
\ \protect \BOthers {.}}{%
Kosugi%
\ \protect \BOthers {.}}{%
{\protect \APACyear {2009}}%
}]{%
Kosugi2009}
\APACinsertmetastar {%
Kosugi2009}%
\begin{APACrefauthors}%
Kosugi, T.%
, Matsuzaki, K.%
, Sakao, R.%
, Bettonvil, F\BPBI C.%
, Sutterlin, P.%
\BCBL {}\ \BBA {} Hammerschlag, R\BPBI H.%
\end{APACrefauthors}%
\unskip\
\newblock
\APACrefYearMonthDay{2009}{}{},
\newblock
\unskip
\newblock
\APACjournalVolNumPages{Sol. Phys.}{243}{}{3}.
\PrintBackRefs{\CurrentBib}

\bibitem [\protect \citeauthoryear {%
Power%
\ \protect \BOthers {.}}{%
Power%
\ \protect \BOthers {.}}{%
{\protect \APACyear {1975}}%
}]{%
Paivio1975}
\APACinsertmetastar {%
Paivio1975}%
\begin{APACrefauthors}%
Power, J\BPBI D.%
, Cohen, A\BPBI L.%
, Nelson, S\BPBI M.%
\ et al.\end{APACrefauthors}%
\unskip\
\newblock
\APACrefYearMonthDay{1975}{}{},
\newblock
\unskip
\newblock
\APACjournalVolNumPages{Cognition}{37}{2}{635}.
\PrintBackRefs{\CurrentBib}

\bibitem [\protect \citeauthoryear {%
Rutten%
}{%
Rutten%
}{%
{\protect \APACyear {2007}}%
}]{%
Rutten2007}
\APACinsertmetastar {%
Rutten2007}%
\begin{APACrefauthors}%
Rutten, R\BPBI J.%
\end{APACrefauthors}%
\unskip\
\newblock
\APACrefYearMonthDay{2007}{}{},
\newblock
{\BBOQ}\APACrefatitle {The Physics of Chromospheric Plasmas} {The Physics of
  Chromospheric Plasmas}.{\BBCQ}
\newblock
\BIn{} P.~Heinzel, I.~Dorotovic\BCBL {}\ \BBA {} R\BPBI J.~Rutten\ (\BEDS),
  \APACrefbtitle {ASP Conf. Ser.} {ASP Conf. Ser.}\ \BVOL~368, \BPG~27.
\PrintBackRefs{\CurrentBib}

\bibitem [\protect \citeauthoryear {%
Stix%
}{%
Stix%
}{%
{\protect \APACyear {2004}}%
}]{%
Stix2004}
\APACinsertmetastar {%
Stix2004}%
\begin{APACrefauthors}%
Stix, M.%
\end{APACrefauthors}%
\unskip\
\newblock
\APACrefYear{2004},
\newblock
\APACrefbtitle {Astronomy and Astrophysics Library} {Astronomy and Astrophysics
  Library}\ (\PrintOrdinal{2}\ \BEd).
\newblock
\APACaddressPublisher{Berlin}{Springer}.
\PrintBackRefs{\CurrentBib}

\bibitem [\protect \citeauthoryear {%
{Strunk Jr.}%
\ \BBA {} White%
}{%
{Strunk Jr.}%
\ \BBA {} White%
}{%
{\protect \APACyear {1979}}%
}]{%
Strunk1979}
\APACinsertmetastar {%
Strunk1979}%
\begin{APACrefauthors}%
{Strunk Jr.}, W.%
\BCBT {}\ \BBA {} White, E\BPBI B.%
\end{APACrefauthors}%
\unskip\
\newblock
\APACrefYear{1979},
\newblock
\APACrefbtitle {The Elements of Style} {The Elements of Style}\
  (\PrintOrdinal{3}\ \BEd).
\newblock
\APACaddressPublisher{New York}{MacMillan}.
\PrintBackRefs{\CurrentBib}

\end{thebibliography}


\begin{thebibliography}{}

\bibitem [\protect \citeauthoryear {%
{Bicknell}%
, {Mukherjee}%
, {Wagner}%
, {Sutherland}%
\BCBL {}\ \BBA {} {Nesvadba}%
}{%
{Bicknell}%
\ \protect \BOthers {.}}{%
{\protect \APACyear {2018}}%
}]{%
2018MNRAS.475.3493B}
\APACinsertmetastar {%
2018MNRAS.475.3493B}%
\begin{APACrefauthors}%
{Bicknell}, G\BPBI V.%
, {Mukherjee}, D.%
, {Wagner}, A\BPBI Y.%
, {Sutherland}, R\BPBI S.%
\BCBL {}\ \BBA {} {Nesvadba}, N\BPBI P\BPBI H.%
\end{APACrefauthors}%
\unskip\
\newblock
\APACrefYearMonthDay{2018}{{\APACmonth{04}}}{},
\newblock
\unskip
\newblock
\APACjournalVolNumPages{\mnras}{475}{3}{3493-3501}.
\PrintBackRefs{\CurrentBib}

\bibitem [\protect \citeauthoryear {%
{Blandford}%
\ \BBA {} {Znajek}%
}{%
{Blandford}%
\ \BBA {} {Znajek}%
}{%
{\protect \APACyear {1977}}%
}]{%
1977MNRAS.179..433B}
\APACinsertmetastar {%
1977MNRAS.179..433B}%
\begin{APACrefauthors}%
{Blandford}, R\BPBI D.%
\BCBT {}\ \BBA {} {Znajek}, R\BPBI L.%
\end{APACrefauthors}%
\unskip\
\newblock
\APACrefYearMonthDay{1977}{{\APACmonth{05}}}{},
\newblock
\unskip
\newblock
\APACjournalVolNumPages{\mnras}{179}{}{433-456}.
\PrintBackRefs{\CurrentBib}

\bibitem [\protect \citeauthoryear {%
{Childs, H. et al.}%
}{%
{Childs, H. et al.}%
}{%
{\protect \APACyear {2012}}%
}]{%
Childs}
\APACinsertmetastar {%
Childs}%
\begin{APACrefauthors}%
{Childs, H. et al.}%
\end{APACrefauthors}%
\unskip\
\newblock
\APACrefYearMonthDay{2012}{}{},
\newblock
\unskip
\newblock
\APACjournalVolNumPages{Proceedings of SciDAC 2011
  (http://press.mcs.anl.gov/scidac2011)}{}{}{}.
\PrintBackRefs{\CurrentBib}

\bibitem [\protect \citeauthoryear {%
{Holt}%
, {Tadhunter}%
, {Morganti}%
\BCBL {}\ \BBA {} {Emonts}%
}{%
{Holt}%
\ \protect \BOthers {.}}{%
{\protect \APACyear {2011}}%
}]{%
Holt11}
\APACinsertmetastar {%
Holt11}%
\begin{APACrefauthors}%
{Holt}, J.%
, {Tadhunter}, C\BPBI N.%
, {Morganti}, R.%
\BCBL {}\ \BBA {} {Emonts}, B\BPBI H\BPBI C.%
\end{APACrefauthors}%
\unskip\
\newblock
\APACrefYearMonthDay{2011}{{\APACmonth{01}}}{},
\newblock
\unskip
\newblock
\APACjournalVolNumPages{\mnras}{410}{3}{1527-1536}.
\PrintBackRefs{\CurrentBib}

\bibitem [\protect \citeauthoryear {%
{Labiano, A. et al.}%
}{%
{Labiano, A. et al.}%
}{%
{\protect \APACyear {2005}}%
}]{%
labiano05}
\APACinsertmetastar {%
labiano05}%
\begin{APACrefauthors}%
{Labiano, A. et al.}%
\end{APACrefauthors}%
\unskip\
\newblock
\APACrefYearMonthDay{2005}{{\APACmonth{06}}}{},
\newblock
\unskip
\newblock
\APACjournalVolNumPages{\aap}{436}{}{493-501}.
\PrintBackRefs{\CurrentBib}

\bibitem [\protect \citeauthoryear {%
{Morganti}%
\ \BBA {} {Oosterloo}%
}{%
{Morganti}%
\ \BBA {} {Oosterloo}%
}{%
{\protect \APACyear {2018}}%
}]{%
2018A&ARv..26....4M}
\APACinsertmetastar {%
2018A&ARv..26....4M}%
\begin{APACrefauthors}%
{Morganti}, R.%
\BCBT {}\ \BBA {} {Oosterloo}, T.%
\end{APACrefauthors}%
\unskip\
\newblock
\APACrefYearMonthDay{2018}{{\APACmonth{07}}}{},
\newblock
\unskip
\newblock
\APACjournalVolNumPages{\aapr}{26}{1}{4}.
\PrintBackRefs{\CurrentBib}

\bibitem [\protect \citeauthoryear {%
{Morganti}%
, {Veilleux}%
, {Oosterloo}%
, {Teng}%
\BCBL {}\ \BBA {} {Rupke}%
}{%
{Morganti}%
\ \protect \BOthers {.}}{%
{\protect \APACyear {2016}}%
}]{%
2016A&A...593A..30M}
\APACinsertmetastar {%
2016A&A...593A..30M}%
\begin{APACrefauthors}%
{Morganti}, R.%
, {Veilleux}, S.%
, {Oosterloo}, T.%
, {Teng}, S\BPBI H.%
\BCBL {}\ \BBA {} {Rupke}, D.%
\end{APACrefauthors}%
\unskip\
\newblock
\APACrefYearMonthDay{2016}{{\APACmonth{09}}}{},
\newblock
\unskip
\newblock
\APACjournalVolNumPages{\aap}{593}{}{A30}.
\PrintBackRefs{\CurrentBib}

\bibitem [\protect \citeauthoryear {%
{Mukherjee}%
, {Bicknell}%
, {Sutherland}%
\BCBL {}\ \BBA {} {Wagner}%
}{%
{Mukherjee}%
\ \protect \BOthers {.}}{%
{\protect \APACyear {2016}}%
}]{%
2016MNRAS.461..967M}
\APACinsertmetastar {%
2016MNRAS.461..967M}%
\begin{APACrefauthors}%
{Mukherjee}, D.%
, {Bicknell}, G\BPBI V.%
, {Sutherland}, R.%
\BCBL {}\ \BBA {} {Wagner}, A.%
\end{APACrefauthors}%
\unskip\
\newblock
\APACrefYearMonthDay{2016}{{\APACmonth{09}}}{},
\newblock
\unskip
\newblock
\APACjournalVolNumPages{\mnras}{461}{1}{967-983}.
\PrintBackRefs{\CurrentBib}

\bibitem [\protect \citeauthoryear {%
{O'Dea}%
\ \protect \BOthers {.}}{%
{O'Dea}%
\ \protect \BOthers {.}}{%
{\protect \APACyear {2002}}%
}]{%
ODea02}
\APACinsertmetastar {%
ODea02}%
\begin{APACrefauthors}%
{O'Dea}, C\BPBI P.%
, {de Vries}, W\BPBI H.%
, {Koekemoer}, A\BPBI M.%
\ et al.\end{APACrefauthors}%
\unskip\
\newblock
\APACrefYearMonthDay{2002}{{\APACmonth{05}}}{},
\newblock
\unskip
\newblock
\APACjournalVolNumPages{\aj}{123}{5}{2333-2351}.
\PrintBackRefs{\CurrentBib}

\bibitem [\protect \citeauthoryear {%
{Owsianik}%
\ \BBA {} {Conway}%
}{%
{Owsianik}%
\ \BBA {} {Conway}%
}{%
{\protect \APACyear {1998}}%
}]{%
1998A&A...337...69O}
\APACinsertmetastar {%
1998A&A...337...69O}%
\begin{APACrefauthors}%
{Owsianik}, I.%
\BCBT {}\ \BBA {} {Conway}, J\BPBI E.%
\end{APACrefauthors}%
\unskip\
\newblock
\APACrefYearMonthDay{1998}{{\APACmonth{09}}}{},
\newblock
\unskip
\newblock
\APACjournalVolNumPages{\aap}{337}{}{69-79}.
\PrintBackRefs{\CurrentBib}

\bibitem [\protect \citeauthoryear {%
{Perucho}%
\ \protect \BOthers {.}}{%
{Perucho}%
\ \protect \BOthers {.}}{%
{\protect \APACyear {2010}}%
}]{%
2010A&A...519A..41P}
\APACinsertmetastar {%
2010A&A...519A..41P}%
\begin{APACrefauthors}%
{Perucho}, M.%
, {Mart{\'\i}}, J\BPBI M.%
, {Cela}, J\BPBI M.%
, {Hanasz}, M.%
, {de La Cruz}, R.%
\BCBL {}\ \BBA {} {Rubio}, F.%
\end{APACrefauthors}%
\unskip\
\newblock
\APACrefYearMonthDay{2010}{{\APACmonth{09}}}{},
\newblock
\unskip
\newblock
\APACjournalVolNumPages{\aap}{519}{}{A41}.
\PrintBackRefs{\CurrentBib}

\bibitem [\protect \citeauthoryear {%
{Perucho}%
, {Mart{\'\i}}%
\BCBL {}\ \BBA {} {Quilis}%
}{%
{Perucho}%
\ \protect \BOthers {.}}{%
{\protect \APACyear {2019}}%
}]{%
2019MNRAS.482.3718P}
\APACinsertmetastar {%
2019MNRAS.482.3718P}%
\begin{APACrefauthors}%
{Perucho}, M.%
, {Mart{\'\i}}, J\BHBI M.%
\BCBL {}\ \BBA {} {Quilis}, V.%
\end{APACrefauthors}%
\unskip\
\newblock
\APACrefYearMonthDay{2019}{{\APACmonth{01}}}{},
\newblock
\unskip
\newblock
\APACjournalVolNumPages{\mnras}{482}{3}{3718-3735}.
\PrintBackRefs{\CurrentBib}

\bibitem [\protect \citeauthoryear {%
{Perucho}%
, {Mart{\'\i}}%
, {Quilis}%
\BCBL {}\ \BBA {} {Ricciardelli}%
}{%
{Perucho}%
\ \protect \BOthers {.}}{%
{\protect \APACyear {2014}}%
}]{%
2014MNRAS.445.1462P}
\APACinsertmetastar {%
2014MNRAS.445.1462P}%
\begin{APACrefauthors}%
{Perucho}, M.%
, {Mart{\'\i}}, J\BHBI M.%
, {Quilis}, V.%
\BCBL {}\ \BBA {} {Ricciardelli}, E.%
\end{APACrefauthors}%
\unskip\
\newblock
\APACrefYearMonthDay{2014}{{\APACmonth{12}}}{},
\newblock
\unskip
\newblock
\APACjournalVolNumPages{\mnras}{445}{2}{1462-1481}.
\PrintBackRefs{\CurrentBib}

\bibitem [\protect \citeauthoryear {%
{Perucho}%
, {Quilis}%
\BCBL {}\ \BBA {} {Mart{\'\i}}%
}{%
{Perucho}%
\ \protect \BOthers {.}}{%
{\protect \APACyear {2011}}%
}]{%
2011ApJ...743...42P}
\APACinsertmetastar {%
2011ApJ...743...42P}%
\begin{APACrefauthors}%
{Perucho}, M.%
, {Quilis}, V.%
\BCBL {}\ \BBA {} {Mart{\'\i}}, J\BHBI M.%
\end{APACrefauthors}%
\unskip\
\newblock
\APACrefYearMonthDay{2011}{}{},
\newblock
\unskip
\newblock
\APACjournalVolNumPages{\apj}{743}{1}{42}.
\PrintBackRefs{\CurrentBib}

\bibitem [\protect \citeauthoryear {%
{Polatidis}%
\ \BBA {} {Conway}%
}{%
{Polatidis}%
\ \BBA {} {Conway}%
}{%
{\protect \APACyear {2003}}%
}]{%
2003PASA...20...69P}
\APACinsertmetastar {%
2003PASA...20...69P}%
\begin{APACrefauthors}%
{Polatidis}, A\BPBI G.%
\BCBT {}\ \BBA {} {Conway}, J\BPBI E.%
\end{APACrefauthors}%
\unskip\
\newblock
\APACrefYearMonthDay{2003}{{\APACmonth{01}}}{},
\newblock
\unskip
\newblock
\APACjournalVolNumPages{\pasa}{20}{1}{69-74}.
\PrintBackRefs{\CurrentBib}

\bibitem [\protect \citeauthoryear {%
{Reynaldi}%
\ \BBA {} {Feinstein}%
}{%
{Reynaldi}%
\ \BBA {} {Feinstein}%
}{%
{\protect \APACyear {2016}}%
}]{%
reynaldi16}
\APACinsertmetastar {%
reynaldi16}%
\begin{APACrefauthors}%
{Reynaldi}, V.%
\BCBT {}\ \BBA {} {Feinstein}, C.%
\end{APACrefauthors}%
\unskip\
\newblock
\APACrefYearMonthDay{2016}{{\APACmonth{01}}}{},
\newblock
\unskip
\newblock
\APACjournalVolNumPages{\mnras}{455}{2}{2242}.
\PrintBackRefs{\CurrentBib}

\bibitem [\protect \citeauthoryear {%
{Schulz}%
\ \protect \BOthers {.}}{%
{Schulz}%
\ \protect \BOthers {.}}{%
{\protect \APACyear {2021}}%
}]{%
2021A&A...647A..63S}
\APACinsertmetastar {%
2021A&A...647A..63S}%
\begin{APACrefauthors}%
{Schulz}, R.%
, {Morganti}, R.%
, {Nyland}, K.%
, {Paragi}, Z.%
, {Mahony}, E\BPBI K.%
\BCBL {}\ \BBA {} {Oosterloo}, T.%
\end{APACrefauthors}%
\unskip\
\newblock
\APACrefYearMonthDay{2021}{{\APACmonth{03}}}{},
\newblock
\unskip
\newblock
\APACjournalVolNumPages{\aap}{647}{}{A63}.
\PrintBackRefs{\CurrentBib}

\bibitem [\protect \citeauthoryear {%
{Shih}%
, {Stockton}%
\BCBL {}\ \BBA {} {Kewley}%
}{%
{Shih}%
\ \protect \BOthers {.}}{%
{\protect \APACyear {2013}}%
}]{%
Shih13}
\APACinsertmetastar {%
Shih13}%
\begin{APACrefauthors}%
{Shih}, H\BHBI Y.%
, {Stockton}, A.%
\BCBL {}\ \BBA {} {Kewley}, L.%
\end{APACrefauthors}%
\unskip\
\newblock
\APACrefYearMonthDay{2013}{}{},
\newblock
\unskip
\newblock
\APACjournalVolNumPages{\apj}{772}{2}{138}.
\PrintBackRefs{\CurrentBib}

\bibitem [\protect \citeauthoryear {%
{The Event Horizon Telescope Collaboration}%
}{%
{The Event Horizon Telescope Collaboration}%
}{%
{\protect \APACyear {2019}}%
}]{%
2019ApJ...875L...1E}
\APACinsertmetastar {%
2019ApJ...875L...1E}%
\begin{APACrefauthors}%
{The Event Horizon Telescope Collaboration}.%
\end{APACrefauthors}%
\unskip\
\newblock
\APACrefYearMonthDay{2019}{}{},
\newblock
\unskip
\newblock
\APACjournalVolNumPages{\apjl}{875}{1}{L1}.
\PrintBackRefs{\CurrentBib}

\bibitem [\protect \citeauthoryear {%
{Vaidya}%
, {Mignone}%
, {Bodo}%
\BCBL {}\ \BBA {} {Massaglia}%
}{%
{Vaidya}%
\ \protect \BOthers {.}}{%
{\protect \APACyear {2015}}%
}]{%
2015A&A...580A.110V}
\APACinsertmetastar {%
2015A&A...580A.110V}%
\begin{APACrefauthors}%
{Vaidya}, B.%
, {Mignone}, A.%
, {Bodo}, G.%
\BCBL {}\ \BBA {} {Massaglia}, S.%
\end{APACrefauthors}%
\unskip\
\newblock
\APACrefYearMonthDay{2015}{{\APACmonth{08}}}{},
\newblock
\unskip
\newblock
\APACjournalVolNumPages{\aap}{580}{}{A110}.
\PrintBackRefs{\CurrentBib}

\bibitem [\protect \citeauthoryear {%
{Wagner}%
\ \BBA {} {Bicknell}%
}{%
{Wagner}%
\ \BBA {} {Bicknell}%
}{%
{\protect \APACyear {2011}}%
}]{%
2011ApJ...728...29W}
\APACinsertmetastar {%
2011ApJ...728...29W}%
\begin{APACrefauthors}%
{Wagner}, A\BPBI Y.%
\BCBT {}\ \BBA {} {Bicknell}, G\BPBI V.%
\end{APACrefauthors}%
\unskip\
\newblock
\APACrefYearMonthDay{2011}{{\APACmonth{02}}}{},
\newblock
\unskip
\newblock
\APACjournalVolNumPages{\apj}{728}{1}{29}.
\PrintBackRefs{\CurrentBib}

\bibitem [\protect \citeauthoryear {%
{Wagner}%
, {Bicknell}%
\BCBL {}\ \BBA {} {Umemura}%
}{%
{Wagner}%
\ \protect \BOthers {.}}{%
{\protect \APACyear {2012}}%
}]{%
2012ApJ...757..136W}
\APACinsertmetastar {%
2012ApJ...757..136W}%
\begin{APACrefauthors}%
{Wagner}, A\BPBI Y.%
, {Bicknell}, G\BPBI V.%
\BCBL {}\ \BBA {} {Umemura}, M.%
\end{APACrefauthors}%
\unskip\
\newblock
\APACrefYearMonthDay{2012}{{\APACmonth{10}}}{},
\newblock
\unskip
\newblock
\APACjournalVolNumPages{\apj}{757}{2}{136}.
\PrintBackRefs{\CurrentBib}

\bibitem [\protect \citeauthoryear {%
{Zovaro}%
\ \protect \BOthers {.}}{%
{Zovaro}%
\ \protect \BOthers {.}}{%
{\protect \APACyear {2019}}%
}]{%
2019MNRAS.489.4944Z}
\APACinsertmetastar {%
2019MNRAS.489.4944Z}%
\begin{APACrefauthors}%
{Zovaro}, H\BPBI R\BPBI M.%
, {Nesvadba}, N\BPBI P\BPBI H.%
, {Sharp}, R.%
, {Bicknell}, G\BPBI V.%
, {Groves}, B.%
, {Mukherjee}, D.%
\BCBL {}\ \BBA {} {Wagner}, A\BPBI Y.%
\end{APACrefauthors}%
\unskip\
\newblock
\APACrefYearMonthDay{2019}{{\APACmonth{11}}}{},
\newblock
\unskip
\newblock
\APACjournalVolNumPages{\mnras}{489}{4}{4944-4961}.
\PrintBackRefs{\CurrentBib}

\end{thebibliography}




\begin{biography}{\includegraphics[width=55pt,height=70pt]{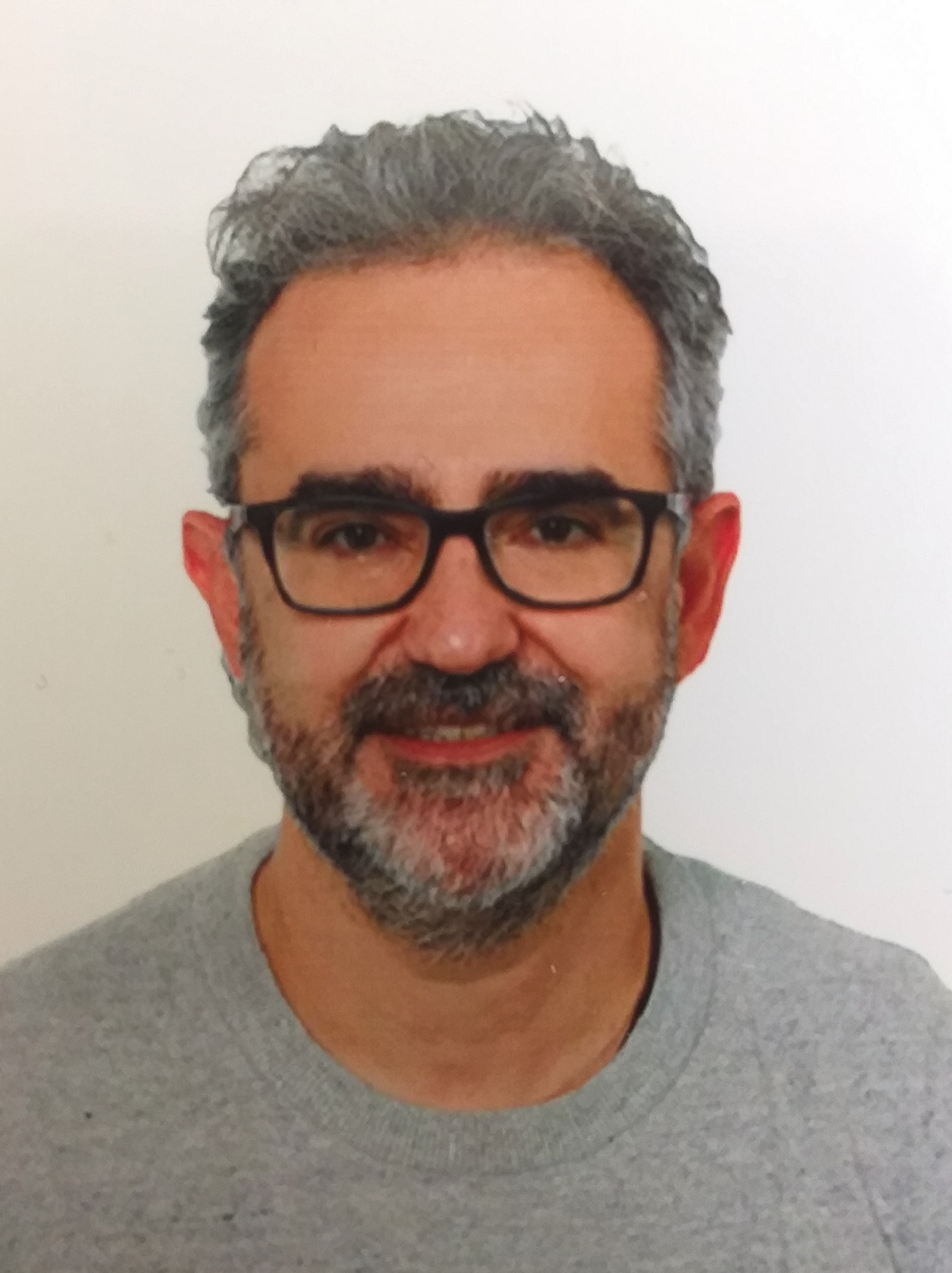}}{\textbf{Manel Perucho.} Manel Perucho i Pla is a Professor at the Facultat de F\'{\i}sica, in the Departament d'Astronomia i Astrof\'{\i}sica of the Universitat de Val\`encia. His research interests are focused in the topics: Relativistic Astrophysics, outflows in AGN, and binary stars.}
\end{biography}

\end{document}